\documentclass[%
 preprint, 
frontmatterverbose, 
 amsmath,amssymb,
 aps, physrev,
]{revtex4-2}

\usepackage{graphicx}
\usepackage{dcolumn}
\usepackage{bm}
\usepackage[dvipsnames]{xcolor}
\usepackage{hyperref}

\newcommand{\wtfrac}{\left(\frac{\sqrt{-g}}{\omega} \right)^{-2 / D}}

\newcommand{\lateralfrac}{\left( \frac{\sqrt{q}}{\omega_{\Gamma}} \right)^{- \frac{2}{D - 1}} }

\newcommand{\be}{\begin{equation}}
\newcommand{\ee}{\end{equation}}

\begin{document}


\title{\textbf{Weyl-transverse gravity with boundaries} 
}%

\author{Gloria Odak}
\affiliation{
Institute of Theoretical Physics, Faculty of Mathematics and Physics, Charles University, Prague,
V Hole\v{s}ovi\v{c}kách 2, 180 00 Prague 8, Czech Republic
}%
\email{gloria.odak@matfyz.cuni.cz}
\author{Salvatore Ribisi}
\affiliation{%
School of Physics and Astronomy, Beijing Normal University, Beijing 100875, China.
}%
\email{salvatore.ribisi@icloud.com}

\date{\today}

\begin{abstract}
We develop the covariant phase space formulation of Weyl–transverse gravity (WTG) in the presence of general timelike and spacelike boundaries. WTG is classically equivalent to General Relativity (GR) but possesses a reduced gauge symmetry consisting of Weyl transformations and transverse diffeomorphisms, together with a fixed background volume form. This structure modifies the variational principle and the definition of conserved quantities relative to GR.

We derive the symplectic potential, presymplectic current, and Hamiltonian generators associated with transverse diffeomorphisms, and we identify a set of boundary conditions under which the WTG action is differentiable. These include Dirichlet and Neumann conditions for both the auxiliary Weyl-invariant metric and the dynamical metric, as well as a natural implementation of York boundary conditions, for which WTG exhibits a particularly transparent geometric formulation.

We obtain the Noether current and surface charge, clarify the role of the Lagrangian ambiguity related to the cosmological constant, and evaluate the Hamiltonian identity on spacetimes containing a bifurcate Killing horizon. The resulting first-law relation shows that variations of the cosmological constant can contribute nontrivially unless additional physical restrictions are imposed.
\end{abstract}

\maketitle
\newpage
\tableofcontents

%
\section{Introduction}
%

Weyl-transverse gravity (WTG) is a metric theory of gravity whose classical equations of motion coincide with those of General Relativity (GR). The main difference between the two lies in the symmetry group of the spacetime metric. While General Relativity is invariant under general diffeomorphisms, Weyl-transverse gravity is invariant under Weyl transformations and transverse diffeomorphisms (breaking the longitudinal diffeomorphisms). As a consequence, the spacetime four-volume $\omega$ appears as a fixed, non-dynamical background field. This feature may remind the reader of \emph{unimodular gravity}, which is in fact WTG with the Weyl gauge symmetry being fixed. In recent years, the theory has received renovated interest \cite{alonso-serrano_noether_2023,alonso-serrano_spacetime_2025,alonso-serrano_new_2020,alonso-serrano_noether_2022}
, for several independent reasons. First, the theory exhibits radiative stability in the cosmological constant sector. Vacuum energy contributions do not gravitate, thereby addressing the technical naturalness aspect of the cosmological constant problem \cite{barcelo_absence_2018,weinberg_cosmological_1989}. Second, when gravitational dynamics are derived from thermodynamic or entropic principles, the resulting equations often take the form of Weyl–transverse gravity rather than fully diffeomorphism-invariant GR \cite{alonso-serrano_spacetime_2025,alonso-serrano_new_2020}. Finally, the breaking of longitudinal diffeomorphisms has been argued to play a useful role in cosmological settings, where it can support effective dissipation mechanisms relevant for current observational tensions \cite{josset_dark_2017,perez_dark_2019}.

An important  property of transverse diffeomorphisms is that all Killing vector fields of a given spacetime metric automatically belong to this subgroup. Indeed, for a Killing vector field $\xi$,
\begin{align}
	\mathcal{L}_{\xi} g_{\mu\nu} = \nabla_{\mu} \xi_{\nu} + \nabla_{\nu} \xi_{\mu} = 0 \quad \Rightarrow \quad g^{\mu\nu} \mathcal{L}_{\xi} g_{\mu\nu} = 2 \nabla_{\mu} \xi^{\mu} = 0 ,
\end{align}
so $\nabla_{\mu}\xi^{\mu}=0$ holds identically. As a result, all standard spacetime symmetries of solutions to Einstein’s equations (such as stationarity or axial symmetry in black-hole spacetimes) are generated by admissible gauge transformations in WTG. This ensures that the Noether charge construction can be applied in a manner closely analogous to GR for physically relevant symmetries.

The Noether charges for Weyl-transverse gravity have already been calculated in \cite{alonso-serrano_noether_2023}. We expand their result by developing a covariant phase space formulation in the presence of boundaries, adopting the framework introduced in \cite{harlow_covariant_2020}. We identify a set of boundary conditions under which the WTG action is differentiable. Recent works have highlighted the relevance of the boundaries in gauge theories \cite{freidel_edge_2020,freidel_extended_2021,ciambelli_asymptotic_2022}, and we want to fill an existing gap in Weyl-transverse gravity.  A further motivation for this analysis is that WTG is particularly well suited to the implementation of York boundary conditions. In York’s prescription, the conformal class of the induced metric and the trace of the extrinsic curvature are held fixed. In WTG, the fundamental dynamical variable is not the metric associated with the spacetime volume form, but a conformally related metric. This is precisely the quantity fixed by York boundary conditions, making WTG a natural framework for their implementation \cite{odak_brown-york_2021,odak_polarizations_2023}. Given the central role of such boundary conditions in both mathematical relativity and gravitational Hamiltonian formulations \cite{anderson_boundary_2008,anderson_quasilocal_2010}, a systematic boundary analysis of WTG is well motivated.

The paper is structured as follows: in section \ref{sec:wtg}, we introduce Weyl-transverse gravity, we find the equations of motions from an action principle, and we show that it is classically equivalent to General Relativity with or without the presence of matter fields. In section \ref{sec:syst}, we compute the presymplectic potential of the theory, and we identify a set of boundary conditions for the theory. These include York boundary conditions, where the dynamical metric $g_{\mu\nu}$ is fixed, as well as the trace of the extrinsic curvature; Dirichlet boundary conditions, where instead of the trace of the extrinsic curvature, the volume at the boundary is fixed; and then the usual Dirichlet and Neumann boundary conditions imposed on the auxiliary metric (which is the effective spacetime metric but is not the fundamental dynamical field in WTG). In section \ref{sec:noether}, we compute the Noether charges for WTG, and evaluate the Hamiltonian on spacetimes containing a bifurcate Killing horizon; we comment on the resulting first-law relation, in which contributions of the cosmological constant are present.

%
\section{Weyl-transverse gravity}\label{sec:wtg}
%

Weyl-transverse gravity is an action theory formulated in terms of a dynamical metric tensor $g_{\mu\nu}$, and a background $D$-form $\bm{\omega} = \omega d^{D} x$. To make the invariance under Weyl transformations explicit, it is useful to introduce the auxiliary metric $\tilde{g}_{\mu\nu}$. Denoting by $g$ the determinant of the tensor $g_{\mu\nu}$, we define
\begin{align} \label{eq:gtilde}
	\tilde{g}_{\mu\nu} &:= \wtfrac g_{\mu\nu} .
\end{align}

Quantities built from the auxiliary metric are by construction Weyl invariant. The covariant derivative associated with the gauge symmetries of the theory is defined by metric compatibility with the auxiliary metric,
\begin{align}\label{eq:metcomp}
	\tilde{\nabla}_{\rho} \tilde{g}_{\mu\nu} &= 0 .
\end{align}

Transverse diffeomorphisms are generated by vector fields $\xi$ that preserve the background volume form. Equivalently, they are defined by the condition
\begin{align}
	0 = \mathcal{L}_{\xi} \omega &= \mathcal{L}_{\xi} \sqrt{- \tilde{g}} = \frac12 \tilde{g}^{\mu\nu} \mathcal{L}_{\xi} \tilde{g}_{\mu\nu} = \tilde{\nabla}_{\mu} \xi^{\mu} = 0 .
\end{align}

Metric compatibility uniquely fixes the connection coefficients to be the Levi–Civita connection of $\tilde g_{\mu\nu}$,
\begin{align}
	\tilde{\Gamma}_{\mu\nu}^{\rho} &= \frac12 \tilde{g}^{\rho\lambda} \left( \partial_{\mu} \tilde{g}_{\lambda\nu} + \partial_{\nu} \tilde{g}_{\mu\lambda} - \partial_{\lambda} \tilde{g}_{\mu\nu} \right) .
\end{align}

From this connection one constructs the Riemann and Ricci tensors in the usual way. 
Throughout the paper, indices are raised and lowered with the auxiliary metric unless stated otherwise; in particular, $g^{\mu\nu}$ always denotes the inverse of the dynamical metric.

Given a $D$-dimensional manifold $M$, and its boundary $\partial M$, the action of Weyl-transverse gravity is given by
\begin{align}
	S_{\mathrm{WTG}} =& \int_{M} L_{G} + \int_{\partial M} \ell_{G}, \\
	L_{G} :=& \frac{\bm{\omega}}{16 \pi G} \tilde{R} ,
\end{align}

with $\ell_{G}$ a boundary Lagrangian (for now arbitrary), $\bm{\omega}$ a background $D$ form, and $\tilde{R}$ the Ricci scalar of the auxiliary metric $\tilde{g}_{\mu\nu}$. Notice that the volume element associated to the auxiliary metric is given by
\begin{align}
	\sqrt{-\tilde{g}} &= \left( \wtfrac \right)^{D/2} \sqrt{- g} = \omega .
\end{align}

The Lagrangian may additionally contain a term
\begin{equation}
L_\lambda = -\frac{\bm{\omega},\lambda}{8\pi G},
\end{equation}
with $\lambda$ constant. Since $\bm{\omega}$ is fixed, this term does not contribute to the equations of motion. Its role will become relevant when discussing conserved charges.

To compute the variation of the action, we will make use of the following standard identities \cite{wald_general_nodate}:
\begin{align}
	\delta \sqrt{- g} &= \frac12 g^{\mu\nu} \delta g_{\mu\nu} \sqrt{-g} , \label{eq:deltaw} \\
	\delta R &= - R^{\mu\nu} \delta g_{\mu\nu} + \nabla^{\mu} \nabla^{\nu} \delta g_{\mu\nu} - \nabla_{\rho} \nabla^{\rho} \left(g^{\mu\nu} \delta g_{\mu\nu} \right) , \label{eq:deltar} \\
	\nabla_{\mu} R^{\mu}_{\ \ \nu} &= \frac12 \nabla_{\nu} R \label{eq:conbia}.
\end{align}

From Eq. \eqref{eq:gtilde}, we can compute the variation of the auxiliary metric
\begin{align}
	\delta \tilde{g}_{\mu\nu} &= \wtfrac \delta g_{\mu\nu} - \frac{2}{D} g_{\mu\nu} \left( \frac{\sqrt{-g}}{\omega} \right)^{-2/D - 1} \frac{\delta \sqrt{-g}}{\omega} \nonumber \\
	&= \wtfrac \left( \delta g_{\mu\nu} - \frac{1}{D} g_{\mu\nu} g^{\alpha\beta} \delta g_{\alpha\beta} \right) , \label{eq:deltagtilde}
\end{align}

where we used Eq. \eqref{eq:deltaw} to go from the first line to the second line. From this equation we know that $\delta \tilde{g}_{\mu\nu}$ is traceless. Keeping in mind that $\bm{\omega}$ is a background field, we can compute the variation of the Lagrangian:
\begin{align}
	\delta L_{G} =& \frac{\bm{\omega} }{16 \pi G} \delta \tilde{R} = \frac{\bm{\omega} }{16 \pi G} \left( - \tilde{R}^{\mu\nu} \delta \tilde{g}_{\mu\nu} + \tilde{\nabla}^{\mu} \tilde{\nabla}^{\nu} \delta \tilde{g}_{\mu\nu} - \tilde{\nabla}_{\rho} \tilde{\nabla}^{\rho} \left( \tilde{g}^{\mu\nu} \delta \tilde{g}_{\mu\nu} \right) \right) \nonumber \\
	=& \frac{\bm{\omega}}{16 \pi G} \wtfrac \left( - \tilde{R}^{\mu\nu} \delta g_{\mu\nu} + \frac{1}{D} \tilde{R}^{\mu\nu} g_{\mu\nu} g^{\alpha\beta} \delta g_{\alpha\beta} \right) \nonumber \\
	&+ \frac{\bm{\omega}}{16 \pi G} \tilde{\nabla}_{\rho} \left( \tilde{g}^{\mu\rho} \tilde{\nabla}^{\nu} \delta \tilde{g}_{\mu\nu} \right) \nonumber \\
	=& E^{\mu\nu}_{G} \delta g_{\mu\nu} + d \Theta_{G} \label{eq:deltaLG},
\end{align}

where we simplified $\tilde{g}^{\mu\nu} \delta \tilde{g}_{\mu\nu} = 0$, and defined
\begin{align}
	E^{\mu\nu}_{G} :=& \frac{\bm{\omega}}{16 \pi G} \wtfrac \left( - \tilde{R}^{\mu\nu} + \frac1D \tilde{R} \tilde{g}^{\mu\nu} \right) , \\
	\Theta_{G} =& \bm{\omega} \cdot \theta_{G}, \qquad \theta_{G}^{\mu} := \frac{1}{16 \pi G} \ \tilde{g}^{\mu\rho} \tilde{\nabla}^{\nu} \delta \tilde{g}_{\rho\nu} \label{eq:thetag} .
\end{align}

Imposing the stationarity of the action gives us the equations of motion $E^{\mu\nu} = 0$, which assuming non-degenerate solutions, i.e. $\sqrt{-g} \ne 0$, implies
\begin{align} \label{eq:tleomv}
	\tilde{R}^{\mu\nu} - \frac{1}{D} \tilde{R} \tilde{g}^{\mu\nu} = 0 .
\end{align}

We can take the divergence of these equations and find
\begin{align}
	0 &= \tilde{\nabla}_{\mu} \left( \tilde{R}^{\mu\nu} - \frac1D \tilde{R} \tilde{g}^{\mu\nu} \right) = \tilde{\nabla}_{\mu} \tilde{R}^{\mu\nu} - \frac{1}{D} \tilde{g}^{\mu\nu} \tilde{R} \nonumber \\
	&=  \frac12 \tilde{\nabla}^{\nu}\tilde{R} - \frac{1}{D} \tilde{g}^{\mu\nu} \tilde{\nabla}_{\mu} \tilde{R} = \frac{D - 2}{2D} \tilde{g}^{\mu\nu} \tilde{\nabla}_{\mu} \tilde{R} = 0,
\end{align}

where we used the contracted Bianchi identity \eqref{eq:conbia} to go from the first line to the second line. Integrating, we find
\begin{align}
	\frac{D - 2}{2D} \tilde{g}^{\mu\nu} \tilde{R} &= \tilde{g}^{\mu\nu} \Lambda .
\end{align}

Subtracting it from Eq. \eqref{eq:tleomv} we finally get the full Einstein's equations
\begin{align}
	\tilde{R}^{\mu\nu} - \frac{1}{2} \tilde{R} \tilde{g}^{\mu\nu} - \Lambda \tilde{g}^{\mu\nu} &= 0 .
\end{align}

\subsection{Coupling to matter}

Let $\psi^{a}$ denote a collection of matter fields minimally coupled to the auxiliary geometry.
We take the matter Lagrangian to be of the form
\be
L_{\psi} = \mathcal{L}_{\psi}(\psi^{a},\partial\psi^{a},\tilde g_{\mu\nu})\,\bm{\omega},
\ee
where $\mathcal{L}_{\psi}$ is a scalar built from the matter fields, their derivatives, and the
auxiliary metric $\tilde g_{\mu\nu}$. Since the matter fields are taken to be inert under Weyl
transformations, $L_{\psi}$ is Weyl invariant by construction. (For multiple species one simply
sums the corresponding contributions.)

We define the stress-energy tensor by variation with respect to the auxiliary metric,
\be
\tilde T^{\mu\nu} := \frac{2}{\omega}\,\frac{\delta S_{\psi}}{\delta \tilde g_{\mu\nu}},
\qquad
S_{\psi}=\int_M L_{\psi}.
\ee
One could equivalently define it by varying with respect to the dynamical metric
$g_{\mu\nu}$. The two definitions are related by the Weyl-invariant field redefinition
\eqref{eq:gtilde}, and we adopt the present convention to match the standard formulation in the
WTG literature.

The variation of the matter Lagrangian can be written as
\be
\delta L_{\psi}
= \frac{\bm{\omega}}{2}\,\tilde T^{\mu\nu}\,\delta\tilde g_{\mu\nu}
+ E_a\,\delta\psi^{a} + \mathrm{d}\Theta_{\psi}.
\ee
Using \eqref{eq:deltagtilde}, we obtain
\be
\delta L_{\psi}
= \frac{\bm{\omega}}{2}\left(\frac{\sqrt{-g}}{\omega}\right)^{-2/D}
\left(\tilde T^{\mu\nu} - \frac{1}{D}\tilde T\,\tilde g^{\mu\nu}\right)\delta g_{\mu\nu}
+ E_a\,\delta\psi^{a} + \mathrm{d}\Theta_{\psi},
\label{eq:deltaLpsi}
\ee
where $\tilde T:=\tilde g_{\mu\nu}\tilde T^{\mu\nu}$.

Combining this with the gravitational variation \eqref{eq:deltaLG}, the variation of the total
Lagrangian $L=L_G+L_{\psi}$ becomes
\be
\delta L
= -\frac{\bm{\omega}}{16\pi G}\left(\frac{\sqrt{-g}}{\omega}\right)^{-2/D}
\left[
\tilde R^{\mu\nu}-\frac{1}{D}\tilde R\,\tilde g^{\mu\nu}
- 8\pi G\left(\tilde T^{\mu\nu}-\frac{1}{D}\tilde T\,\tilde g^{\mu\nu}\right)
\right]\delta g_{\mu\nu}
+ E_a\,\delta\psi^{a}
+ \mathrm{d}(\Theta_G+\Theta_{\psi}).
\ee
Stationarity with respect to $\delta g_{\mu\nu}$ yields the traceless Einstein equations
\be
\tilde R^{\mu\nu}-\frac{1}{D}\tilde R\,\tilde g^{\mu\nu}
= 8\pi G\left(\tilde T^{\mu\nu}-\frac{1}{D}\tilde T\,\tilde g^{\mu\nu}\right).
\label{eq:tlee}
\ee

Taking the divergence of \eqref{eq:tlee} and using the contracted Bianchi identity gives
\be
\tilde g^{\mu\nu}\tilde\nabla_{\mu}
\left(\frac{D-2}{2D}\tilde R+\frac{8\pi G}{D}\tilde T\right)
= 8\pi G\,\tilde\nabla_{\mu}\tilde T^{\mu\nu}.
\ee
In WTG, invariance under transverse diffeomorphisms does not by itself imply the standard
conservation law $\tilde\nabla_{\mu}\tilde T^{\mu\nu}=0$. Following the usual treatment, we
assume the integrability condition
\be
\tilde\nabla_{\mu}\tilde T^{\mu\nu} = \tilde g^{\mu\nu}\tilde\nabla_{\mu}\mathcal{J},
\ee
for some scalar function $\mathcal{J}$. This condition holds, for instance, when the non-conservation is purely longitudinal in the auxiliary geometry. 
Under this assumption we can integrate to obtain
\be
\frac{D-2}{2D}\tilde R+\frac{8\pi G}{D}\tilde T
= 8\pi G\,\mathcal{J}+\Lambda,
\ee
where $\Lambda$ is an integration constant. Substituting back into \eqref{eq:tlee} yields the
full Einstein equations in the auxiliary geometry,
\be
\tilde R^{\mu\nu}-\frac{1}{2}\tilde R\,\tilde g^{\mu\nu}+\Lambda\,\tilde g^{\mu\nu}
= 8\pi G\left(\tilde T^{\mu\nu}-\mathcal{J}\tilde g^{\mu\nu}\right).
\ee
Defining the conserved stress tensor
\be
\tilde T'^{\mu\nu}:=\tilde T^{\mu\nu}-\mathcal{J}\tilde g^{\mu\nu},
\ee
we can rewrite the equations as
\be
\tilde R^{\mu\nu}-\frac{1}{2}\tilde R\,\tilde g^{\mu\nu}+\Lambda\,\tilde g^{\mu\nu}
= 8\pi G\,\tilde T'^{\mu\nu}.
\label{eq:einsteq}
\ee

%
\section{Symplectic structure}\label{sec:syst}
%

A well-posed variational principle requires more than imposing the equations of motion in the bulk,
one must also control the boundary variation of the action. In the covariant phase space approach,
bulk solutions subject to admissible boundary conditions define the pre-phase space, and the choice
of boundary terms and boundary conditions determines the symplectic structure \cite{harlow_covariant_2020}.
Consider an action of the form

\begin{align}
	S &= \int_{M} L + \int_{\partial M} \ell,
\end{align}

with $M$ a differentiable $D$-dimensional manifold and $\partial M$ its boundary. Its variation is given by
\begin{align}
	\delta S &= \int_{M} \delta L + \int_{\partial M} \delta \ell \nonumber \\
	&= \int_{M} \left( E^{\mu\nu} \delta g_{\mu\nu} + E_{a} \delta \psi^{a} + d \Theta \right) + \int_{\partial M} \delta \ell \nonumber \\
	&= \int_{M} \left( E^{\mu\nu} \delta g_{\mu\nu} + E_{a} \delta \psi^{a} \right) + \int_{\partial M} \left(\Theta + \delta \ell \right) ,
\end{align}

where we used Stokes theorem in the last step. We decompose the boundary into a timelike (lateral) component $\Gamma$ and spacelike components
$\Sigma_\pm$ (initial/final hypersurfaces), so that $\partial M = \Gamma \cup \Sigma_+ \cup \Sigma_-$.
Variations on $\Sigma_\pm$ correspond to variations of the physical state, whereas boundary conditions
are imposed so as to ensure stationarity on $\Gamma$ \cite{harlow_covariant_2020}. We assume the corner
$\partial\Sigma:=\Sigma\cap\Gamma$ is compact, so that integrals of total derivatives on $\partial\Sigma$
vanish. With these conventions,
\begin{align}
	\delta S &= \int_{M} \left( E^{\mu\nu} \delta g_{\mu\nu} + E_{a} \delta \psi^{a} \right) + \int_{\Sigma_{+} - \Sigma_{-}} \left( \Theta + \delta \ell \right) + \int_{\Gamma} \left( \Theta + \delta \ell \right) .
\end{align}

A sufficient condition for the lateral boundary contribution to vanish is
\begin{align}
	 \left.\left( \Theta + \delta \ell \right) \right|_{\Gamma} &= d C , \label{eq:lateralvanish}
\end{align}

where $C$ is a local $\left(D- 2\right)$ form on $\Gamma$ constructed out of the dynamical and background fields, their variations, and derivatives of both. The variation of the action then takes the form
\begin{align}
	\delta S &= \int_{M} \left( E^{\mu\nu} \delta g_{\mu\nu} + E_{a} \delta \psi^{a} \right) + \int_{\Sigma_{+} - \Sigma_{-}} \left( \Theta + \delta \ell \right) + \int_{\partial \Gamma} C \nonumber \\
	&= \int_{M} \left( E^{\mu\nu} \delta g_{\mu\nu} + E_{a} \delta \psi^{a} \right) + \int_{\Sigma_{+} - \Sigma_{-}} \left( \Theta + \delta \ell - d C \right) .
\end{align}

The field configurations satisfying Eq. \eqref{eq:lateralvanish} define the configuration space $\mathcal{C}$. 
On $\mathcal{C}$, $\Theta$ and $C$ can be viewed as one-forms, and $\delta$ as the exterior derivative
in the exterior algebra of $\mathcal C$. We define the $(D-1)$-form integrated over Cauchy surfaces,
\begin{align}\label{eq:cauchypsi}
	\Psi := \Theta + \delta \ell - d C .
\end{align}

In our case, the boundary condition \eqref{eq:lateralvanish} takes the form
\begin{align}
	\left. \left(\Theta_{G} + \Theta_{\psi} + \delta \ell_{G} + \delta \ell_{\psi} \right) \right|_{\Gamma} &= d C .
\end{align}

Since the matter contribution depends on the specific choice of fields and boundary terms, we exploit
linearity and decompose $C=C_G+C_\psi$, requiring separately
\begin{align}
	\left. \left(\Theta_{G} + \delta \ell_{G}\right) \right|_{\Gamma} &= d C_{G} , \label{eq:thdcg}\\
	\left. \left(\Theta_{\psi} + \delta \ell_{\psi}\right) \right|_{\Gamma} &= d C_{\psi} .
\end{align}

This decomposition is always possible because of our freedom in the definition of the boundary Lagrangians $\ell$. We now focus on the gravitational sector and compute the pullback of $\Theta_G$ to $\partial M$.

\subsection{Variation of the auxiliary metric}

Let $\tilde n^\mu$ be the unit normal to $\partial M$ with respect to the auxiliary metric,
normalized as
\be
\tilde n^\mu \tilde n^\nu \tilde g_{\mu\nu} = s,
\qquad s=\pm 1.
\ee
With this choice, both $\tilde n^\mu$ and the induced volume form
$\bm{\omega}_{\partial M}:=\iota_{\tilde n}\bm{\omega}$ are Weyl invariant. The induced metric and
extrinsic curvature (built from $\tilde g_{\mu\nu}$) are
\begin{align}
	\tilde{K}^{\mu\nu} &= \tilde{q}^{\mu\lambda} \tilde{\nabla}_{\lambda} \tilde{n}^{\nu},
\end{align}

and we note its trace as $\tilde{K} = \tilde{K}^{\mu\nu} \tilde{q}_{\mu\nu} = \tilde{K}^{\mu\nu} \tilde{g}_{\mu\nu}$. Its variation can be computed (see for instance \cite{wald_general_nodate}) and is given by
\begin{align}
	\delta \tilde{K} &= - \frac12 \tilde{K}^{\mu\nu} \delta \tilde{g}_{\mu\nu} + \frac12 \tilde{g}^{\mu\nu} \tilde{n}^{\lambda}\tilde \nabla_{\lambda} \delta \tilde{g}_{\mu\nu} - \frac12 \tilde{n}^{\alpha} \tilde{\nabla}^{\beta} \delta \tilde{g}_{\alpha\beta} - \frac12 \tilde{D}_{\mu} \left(\tilde{q}^{\mu\nu} \tilde{n}^{\alpha} \delta \tilde{g}_{\nu\alpha} \right) \label{eq:deltak} .
\end{align}

From Eq. \eqref{eq:thetag},
\begin{align}
	\left. \Theta_{G} \right|_{\partial M} =& \frac{\bm{\omega}_{\partial M}}{16 \pi G} \tilde{n}^{\mu}  \tilde{\nabla}^{\nu} \delta \tilde{g}_{\mu\nu} \nonumber \\
	=& \frac{\bm{\omega}_{\partial M}}{16 \pi G} \left( - 2 \delta \tilde{K} - \tilde{K}^{\mu\nu} \delta \tilde{g}_{\mu\nu} - \tilde{D}_{\mu} \left( \tilde{q}^{\mu\nu} \tilde{n}^{\alpha} \delta \tilde{g}_{\nu\alpha} \right) \right) \nonumber \\
	=& - \delta \left( \frac{\bm{\omega}_{\partial M} \tilde{K} }{8 \pi G} \right) + \frac{\tilde{K}}{8 \pi G} \delta \bm{\omega}_{\partial M} + \frac{\bm{\omega}_{\partial M}}{16 \pi G} \left( - \tilde{K}^{\mu\nu} \delta \tilde{g}_{\mu\nu} - \tilde{D}_{\mu} \left( \tilde{q}^{\mu\nu} \tilde{n}^{\alpha} \delta \tilde{g}_{\nu\alpha} \right) \right) \nonumber \\
	=& - \delta \left( \frac{\bm{\omega}_{\partial M} \tilde{K}}{8 \pi G} \right) + \frac{\bm{\omega}_{\partial M}}{16 \pi G} \left( - \tilde{K}^{\mu\nu} + \tilde{K} \tilde{q}^{\mu\nu} \right) \delta \tilde{q}_{\mu\nu} + d C ,\label{eq:thone}
\end{align}

with
\begin{align}
	C =& \bm{\omega}_{\partial M} \cdot c, \qquad \qquad c^{\mu} :=  - \frac{1}{16 \pi G} \tilde{q}^{\mu\nu} \tilde{n}^{\alpha} \delta \tilde{g}_{\nu\alpha} .
\end{align}

Equation \eqref{eq:thone} is formally analogous to the GR expression. Choosing the boundary Lagrangian
\begin{align}
	\ell_{D} &= \frac{\bm{\omega}_{\partial M}}{8 \pi G} \tilde{K},
\end{align}

condition \eqref{eq:thdcg} is satisfied if
\begin{align}
	\left. \left( - \tilde{K}^{\mu\nu} + \tilde{K} \tilde{q}^{\mu\nu} \right) \delta \tilde{g}_{\mu\nu} \right|_{\Gamma} &= 0,
\end{align}

which is compatible with Dirichlet boundary conditions requiring the pullback of $\tilde{g}_{\mu\nu}$ on $\Gamma$ to be fixed. Namely
\begin{align}
	\left. \tilde{q}_{\mu}^{\ \alpha} \tilde{q}_{\nu}^{\ \beta} \delta \tilde{g}_{\alpha\beta} \right|_{\Gamma} = \left. \tilde{q}_{\mu}^{\ \alpha} \tilde{q}_{\nu}^{\ \beta} \delta \tilde{q}_{\alpha\beta} \right|_{\Gamma} = 0 .
\end{align}

It is convenient to introduce the corresponding gravitational momentum
\begin{align}
	\tilde{\Pi}^{\mu\nu} :=& \frac{\bm{\omega}_{\partial M}}{16 \pi G} \left( - \tilde{K}^{\mu\nu} + \tilde{K} \tilde{q}^{\mu\nu} \right) .
\end{align}

Integrating by parts in field space (i.e. in the $\delta$-exterior derivative) yields an equivalent form
adapted to Neumann-type boundary conditions
\begin{align}
	\left. \Theta_{G} \right|_{\partial M} &= - \delta \left( \frac{\bm{\omega}_{\partial M} \tilde{K}}{8 \pi G} \right) + \delta \left( \tilde{\Pi}^{\mu\nu} \tilde{q}_{\mu\nu} \right) - \tilde{q}_{\mu\nu} \delta \tilde{\Pi}^{\mu\nu} + d C \nonumber \\
	&= \left(D - 4\right) \delta \left(\frac{\bm{\omega}_{\partial M} \tilde{K}}{16 \pi G} \right) - \tilde{q}_{\mu\nu} \delta \tilde{\Pi}^{\mu\nu} + d C .
\end{align}

In this case the boundary Lagrangian would be $\ell_{N} = (D - 4)/2 \ell_{D}$. 

\subsection{Variation of the dynamical metric}

The boundary conditions discussed so far were imposed on the auxiliary geometry, i.e. on the pullback
$\tilde q_{\mu\nu}$ of $\tilde g_{\mu\nu}$ to $\partial M$. We now analyze boundary conditions formulated
directly in terms of the pullback $q_{\mu\nu}$ of the dynamical metric $g_{\mu\nu}$. This is the natural
setting to compare with York-type data in GR. On $\partial M$ the auxiliary and dynamical induced metrics are related by a conformal rescaling.
Writing the relation in terms of the boundary determinants,
\begin{align}
	\tilde{q}_{\mu\nu} &= \wtfrac q_{\mu\nu} = \left( \frac{\sqrt{sq}}{\omega_{\partial M}} \right)^{- \frac{2}{D - 1}} q_{\mu\nu},
\end{align}

with $s$ determining the orientation of the boundary, $\tilde{n}^{\mu} \tilde{n}^{\nu} \tilde{g}_{\mu\nu} = s$. Its variation is slightly more complicated than the variation of the auxiliary metric, because $\omega_{\partial M}$ is now allowed to vary. In what follows we focus on the lateral boundary $\Gamma$ and set $s=+1$ for simplicity. Allowing the boundary volume form $\omega_\Gamma$ to vary, the variation of $\tilde q_{\mu\nu}$ reads

\be
\delta\tilde q_{\mu\nu}
=
\left(\frac{\sqrt{q}}{\omega_{\Gamma}}\right)^{-2/(D-1)}
\left(
\delta q_{\mu\nu}
-\frac{1}{D-1}\,q_{\mu\nu}\,q^{\alpha\beta}\delta q_{\alpha\beta}
+\frac{2}{D-1}\,q_{\mu\nu}\,\frac{\delta\omega_{\Gamma}}{\omega_{\Gamma}}
\right).
\ee

It is convenient to introduce the traceless variation of the induced metric,
\be
\delta q_{\langle\mu\nu\rangle}
:= \delta q_{\mu\nu}-\frac{1}{D-1}\,q_{\mu\nu}\,q^{\alpha\beta}\delta q_{\alpha\beta},
\ee
so that $\delta\tilde q_{\mu\nu}$ decomposes into a traceless part (controlled by $\delta q_{\langle\mu\nu\rangle}$)
and a pure-trace part (controlled by $\delta\omega_\Gamma$).

Substituting this expression into \eqref{eq:thone} gives, on $\Gamma$,
\be
\Theta_G\big|_{\Gamma}
= -\delta\ell_D + \mathrm{d}C
+ \Pi^{\mu\nu}\,\delta q_{\mu\nu}
+ \frac{D-2}{D-1}\,\frac{1}{8\pi G}\,\tilde K\,\delta\bm{\omega}_{\Gamma},
\label{eq:dyndir_clean}
\ee
where $\ell_D=\frac{\bm{\omega}_\Gamma}{8\pi G}\tilde K$ is the auxiliary Dirichlet boundary term and we defined
the dynamical-metric momentum
\be
\Pi^{\mu\nu}
:= \left(\frac{\sqrt{q}}{\omega_{\Gamma}}\right)^{-2/(D-1)}
\frac{\bm{\omega}_{\Gamma}}{16\pi G}\left(-\tilde K^{\mu\nu}+\frac{1}{D-1}\tilde K\,\tilde q^{\mu\nu}\right).
\ee
The key point is that $\Pi^{\mu\nu}$ is traceless with respect to $q_{\mu\nu}$, and therefore couples only to the
traceless part $\delta q_{\langle\mu\nu\rangle}$. As a consequence, fixing $q_{\mu\nu}$ on $\Gamma$ requires, in addition,
one extra scalar boundary condition controlling the remaining trace/volume variation. Two natural choices are:
(i) fixing the boundary volume form $\bm{\omega}_\Gamma$ (Dirichlet-type), or
(ii) fixing the mean curvature $\tilde K$ (York-type).

To obtain the York-type formulation, we integrate by parts in field space the $\tilde K\,\delta\bm{\omega}_\Gamma$ term in
\eqref{eq:dyndir_clean}. Using $\delta(\tilde K\bm{\omega}_\Gamma)=\tilde K\,\delta\bm{\omega}_\Gamma+\bm{\omega}_\Gamma\,\delta\tilde K$,
we can rewrite \eqref{eq:dyndir_clean} as
\be
\Theta_G\big|_{\Gamma}
= -\delta\ell_Y + \mathrm{d}C
+ \Pi^{\mu\nu}\,\delta q_{\mu\nu}
- \frac{D-2}{D-1}\,\frac{\bm{\omega}_\Gamma}{8\pi G}\,\delta\tilde K,
\label{eq:dyn_york_form}
\ee
with York-type boundary Lagrangian
\be
\ell_Y := \frac{1}{D-1}\,\ell_D
= \frac{1}{D-1}\,\frac{\bm{\omega}_\Gamma}{8\pi G}\tilde K.
\ee

\subsection{Boundary conditions}

We summarize the admissible gravitational boundary terms and the corresponding boundary conditions.
We consider the one-parameter family of boundary Lagrangians
\begin{align}
	\ell_{b} &= \frac{b}{16 \pi G} \tilde{K} \bm{\omega}_{\Gamma} ,
\end{align}

and recall the definitions of the two gravitational momenta: 
\begin{align}
	\tilde{\Pi}^{\mu\nu} :=& - \frac{\bm{\omega}_{\Gamma}}{16 \pi G} \left( \tilde{K}^{\mu\nu} - \tilde{K} \tilde{q}^{\mu\nu} \right) , \\
	\Pi^{\mu\nu} := & - \frac{\bm{\omega}_{\Gamma}}{16 \pi G} \lateralfrac \left( \tilde{K}^{\mu\nu} - \frac{1}{D - 1} \tilde{K} \tilde{q}^{\mu\nu} \right) .
\end{align}

\paragraph{ Auxiliary Dirichlet boundary conditions.}
Using \eqref{eq:thone}, choosing $b=2$ (i.e. $\ell_2=\ell_D=\frac{\bm{\omega}_\Gamma}{8\pi G}\tilde K$) yields
\be
\left.(\Theta_G+\delta\ell_2)\right|_\Gamma
= \tilde\Pi^{\mu\nu}\,\delta\tilde q_{\mu\nu} + \mathrm{d}C,
\qquad
\tilde\Pi^{\mu\nu}
:= \frac{\bm{\omega}_\Gamma}{16\pi G}\left(-\tilde K^{\mu\nu}+\tilde K\,\tilde q^{\mu\nu}\right).
\ee
Stationarity is ensured by fixing the induced auxiliary metric on $\Gamma$,
\be
\left.\delta\tilde q_{\mu\nu}\right|_\Gamma = 0.
\ee
\paragraph{Auxiliary Neumann boundary conditions.}
Rewriting the pullback potential in the Neumann form obtained above, choosing $b=4-D$ gives
\be
\left.(\Theta_G+\delta\ell_{4-D})\right|_\Gamma
= -\tilde q_{\mu\nu}\,\delta\tilde\Pi^{\mu\nu} + \mathrm{d}C,
\qquad
\left.\delta\tilde\Pi^{\mu\nu}\right|_\Gamma = 0.
\ee
\paragraph{Dirichlet boundary conditions.}
From \eqref{eq:dyndir_clean}, choosing again $b=2$ yields
\be
\left.(\Theta_G+\delta\ell_2)\right|_\Gamma
= \Pi^{\mu\nu}\,\delta q_{\mu\nu}
+ \frac{D-2}{D-1}\,\frac{1}{8\pi G}\,\tilde K\,\delta\bm{\omega}_\Gamma
+ \mathrm{d}C.
\ee
Since $\Pi^{\mu\nu}$ is traceless, the natural Dirichlet-type condition fixes the conformal class of $q_{\mu\nu}$
together with the boundary volume form,
\be
\left.\delta q_{\langle\mu\nu\rangle}\right|_\Gamma = 0,
\qquad
\left.\delta\bm{\omega}_\Gamma\right|_\Gamma = 0.
\ee
\paragraph{York boundary conditions.}
From \eqref{eq:dyn_york_form}, choosing $b=\frac{2}{D-1}$ (i.e. $\ell_{2/(D-1)}=\ell_Y$) gives
\be
\left.(\Theta_G+\delta\ell_{2/(D-1)})\right|_\Gamma
= \Pi^{\mu\nu}\,\delta q_{\mu\nu}
- \frac{D-2}{D-1}\,\frac{\bm{\omega}_\Gamma}{8\pi G}\,\delta\tilde K
+ \mathrm{d}C.
\ee
Stationarity is ensured by fixing the conformal class of the induced metric together with the mean curvature,
\be
\left.\delta q_{\langle\mu\nu\rangle}\right|_\Gamma = 0,
\qquad
\left.\delta\tilde K\right|_\Gamma = 0.
\ee

%
\section{Noether current}\label{sec:noether}
%

Given a vector field $\xi$ generating an infinitesimal gauge transformation, we define the Noether current
\begin{align}
	J_{\xi} :=& X_{\xi} \cdot \Theta - \xi \cdot L .
\end{align}
where $X_\xi$ is the field-space vector corresponding to the transformation $\delta_\xi$ and $\Theta$ is the
symplectic potential. The Noether current is a $(D-1)$-form on spacetime and a function on field space.
If $L$ is covariant under $\xi$, then $J_{\xi}$ is closed as a spacetime form:
\begin{align}
	d J_{\xi} &= d \left( X_{\xi} \cdot \Theta \right) - d \left( \xi \cdot L \right) \nonumber\\
	&= X_{\xi} \cdot d \Theta - \mathcal{L}_{\xi} L \nonumber \\
	&= X_{\xi} \cdot \delta L - X_{\xi} \cdot \left( E_{a} \delta \phi^{a} \right) - \mathcal{L}_{\xi} L \nonumber \\
	&= \left( \delta_{\xi} - \mathcal{L}_{\xi} \right) L - E_{a} \delta_{\xi} \phi^{a} = 0.
\end{align}
Closedness does not imply global exactness, nevertheless, as in GR,
one can isolate an exact piece together with terms proportional to the equations of motion. The novelty in WTG is
that (i) $\xi$ is restricted to be transverse and (ii) the presence of the background volume form allows an
additional contribution proportional to $\xi\cdot\bm{\omega}$.
The current associated with the matter fields is
\begin{align}
	J_{\xi}^{\psi} := X_{\xi} \cdot \Theta_{\psi} - \xi \cdot L_{\psi} .
\end{align}
It can be shown to be proportional to the conserved stress-energy tensor $\tilde{T}'^{\mu\nu}$.
In particular, since only the traceless part of the stress-energy tensor enters the gravitational
equations, shifts proportional to $\tilde{g}^{\mu\nu}$ can be absorbed into the scalar $\mathcal{J}$.
We take Eq.~\eqref{eq:deltaLpsi} and contract it with $- X_{\xi}$, obtaining
 \begin{align}
	- X_{\xi} \cdot \delta L_{\psi}
	&= - \frac{\bm{\omega}}{2} \tilde{T}^{\mu\nu} \delta_{\xi} \tilde{g}_{\mu\nu}
	- E_{a} \delta_{\xi} \psi^{a}
	- X_{\xi} \cdot d \Theta_{\psi} \nonumber \\
	- \mathcal{L}_{\xi} L_{\psi}
	&= - \bm{\omega} \tilde{T}^{\mu\nu} \tilde{\nabla}_{(\mu} \xi_{\nu)}
	- E_{a} \mathcal{L}_{\xi} \psi^{a}
	- d \left( X_{\xi} \cdot \Theta_{\psi} \right) .
\end{align}
Combining these expressions gives, on shell,
\begin{align}
	d \left( X_{\xi} \cdot \Theta_{\psi} - \xi \cdot L_{\psi} \right)
	&\approx - \bm{\omega} \tilde{T}^{\mu\nu} \tilde{\nabla}_{\mu} \xi_{\nu} .
\end{align}
Integrating by parts yields
\begin{align}
	- \bm{\omega} \tilde{T}^{\mu\nu} \tilde{\nabla}_{\mu} \xi_{\nu}
	&= - \bm{\omega} \tilde{\nabla}_{\mu} \left( \tilde{T}^{\mu\nu} \xi_{\nu} \right)
	+ \bm{\omega} \xi_{\nu} \tilde{\nabla}_{\mu} \tilde{T}^{\mu\nu} \nonumber \\
	&= - \bm{\omega} \tilde{\nabla}_{\mu} \left( \tilde{T}^{\mu\nu} \xi_{\nu} \right)
	+ \bm{\omega} \xi_{\nu} \tilde{g}^{\mu\nu} \tilde{\nabla}_{\mu} \mathcal{J} \nonumber \\
	&= - \bm{\omega} \tilde{\nabla}_{\mu}
	\left( \left( \tilde{T}^{\mu\nu} - \mathcal{J} \tilde{g}^{\mu\nu} \right) \xi_{\nu} \right)
	- \bm{\omega} \mathcal{J} \tilde{\nabla}_{\mu} \xi^{\mu} \nonumber \\
	&= - \bm{\omega} \tilde{\nabla}_{\mu} \left( \tilde{T}'^{\mu\nu} \xi_{\nu} \right) ,
\end{align}
where in the last step we used the transversality condition $\tilde{\nabla}_{\mu} \xi^{\mu}=0$.
We thus conclude that
\begin{align}
	J_{\xi}^{\psi}
	= \bm{\omega} \cdot j_{\psi},
	\qquad
	j^{\mu}_{\psi} = - \tilde{T}'^{\mu\nu} \xi_{\nu} .
\end{align}

Regarding the gravitational contribution, we find
\begin{align}
	\xi \cdot L_{G} &= \frac{\bm{\omega}_{\mu}}{16 \pi G} \tilde{R} \xi^{\mu} \\
	X_{\xi} \cdot \Theta_{G} &= \frac{\bm{\omega}_{\mu}}{16 \pi G} \tilde{g}^{\alpha \mu} \tilde{\nabla}^{\beta} \delta_{\xi} \tilde{g}_{\alpha\beta} = \frac{\bm{\omega}_{\mu}}{8 \pi G} \tilde{g}^{\alpha\mu} \tilde{\nabla}^{\beta} \tilde{\nabla}_{\left( \alpha \right.} \xi_{\left. \beta \right)} \nonumber \\
	&= \frac{\bm{\omega}_{\mu}}{16 \pi G} \left( \tilde{\nabla}_{\beta} \tilde{\nabla}^{\beta} \xi^{\mu} - \tilde{\nabla}_{\beta} \tilde{\nabla}^{\mu} \xi^{\beta} + 2 \tilde{g}^{\alpha\mu} \tilde{\nabla}_{\beta} \tilde{\nabla}_{\alpha} \xi^{\beta}  - 2 \tilde{g}^{\alpha\mu} \tilde{\nabla}_{\alpha} \tilde{\nabla}_{\beta} \xi^{\beta}  \right) \nonumber \\
	&= \frac{\bm{\omega}_{\mu}}{8 \pi G} \left( \tilde{\nabla}_{\nu} \tilde{\nabla}^{\left[ \nu \right.} \xi^{\left. \mu \right]} + \tilde{g}^{\alpha\mu} \tilde{R}^{\beta}_{\ \alpha \beta \gamma} \xi^{\gamma} \right) = \frac{\bm{\omega}_{\mu}}{8 \pi G} \left( \tilde{\nabla}_{\nu} \tilde{\nabla}^{\left[ \nu \right.} \xi^{\left. \mu \right]} + \tilde{R}^{\mu\nu} \xi_{\nu} \right) ,
\end{align}

where in the last step we used standard commutator identities together with the transversality condition $\tilde{\nabla}_{\mu} \xi^{\mu} = 0$. 

Recalling the Lagrangian ambiguity
\begin{align}
	L_{\lambda} = - \frac{\lambda}{8 \pi G} \bm{\omega},
\end{align}
we obtain
\begin{align}
	J_{\xi} &= X_{\xi} \cdot \Theta_{G} - \xi \cdot L_{G} - \xi \cdot L_{\lambda} + J_{\xi}^{\psi} , \\
	J_{\xi} &= \bm{\omega} \cdot j_{\xi} , \\ 
	j^{\mu}_{\xi} &= \frac{1}{8 \pi G} \left( \tilde{R}^{\mu\nu} - \frac{1}{2} \tilde{R} \tilde{g}^{\mu\nu} - 8 \pi G \tilde{T}'^{\mu\nu} - 8 \pi G \lambda \right) \xi_{\nu} + \frac{1}{8 \pi G} \tilde{\nabla}_{\nu} \tilde{\nabla}^{\left[ \nu \right.} \xi^{\left. \mu \right]} .
\end{align}
Using the equations of motion \eqref{eq:einsteq},  this simplifies to
\begin{align}
	j_{\xi}^{\mu} &\approx \frac{1}{8 \pi G} \left(\Lambda - 8 \pi G \lambda \right) \xi^{\mu} + \frac{1}{8 \pi G} \tilde{\nabla}_{\nu} \tilde{\nabla}^{\left[ \nu \right.} \xi^{\left. \mu \right]} .
\end{align}
The divergence of this expression vanishes identically due to transversality of $\xi$ and antisymmetry
of the second term. Recognizing the latter as the divergence of the GR Noether charge,
\begin{align}
	Q_{\mathrm{GR}} &= - \frac{1}{16 \pi G} \star d \xi ,
\end{align}
we can finally write
\begin{align}
	J_{\xi}
	&\approx d Q_{\mathrm{GR}}
	- \frac{\Lambda - \lambda}{8 \pi G} \, \xi \cdot \bm{\omega} .
\end{align}

\subsection{Hamiltonian flow}

Let us define the presymplectic current $\tilde{\omega}$ as the pullback on pre-phase space $\tilde{\mathcal{P}}$ of $\delta \Psi$, defined in Eq. \eqref{eq:cauchypsi}
\begin{align}
	\tilde{\omega} &= \left.\delta \left( \Theta - d C \right) \right|_{\tilde{\mathcal{P}}}.
\end{align}

It is a closed two-form in pre-phase space, as the exterior derivative and the pullback are commuting operations. It is a closed form in spacetime too, as
\begin{align}
	d \tilde \omega &= \left. \delta d \Theta  \right|_{\tilde{\mathcal{P}}} = \left. \delta \left( \delta L - E_{a} \delta \phi^{a} \right) \right|_{\tilde{\mathcal{P}}} = 0 ,
\end{align}

and it vanishes on the lateral boundary
\begin{align}
	\left. \tilde{\omega} \right|_{\Gamma} &= \left. \delta \left( \Theta + \delta \ell - d C \right) \right|_{\Gamma, \tilde{\mathcal{P}}} = \left. \delta \left(d C - d C \right) \right|_{\tilde{\mathcal{P}}} = 0 .
\end{align}

Hence, the integral of $\tilde{\omega}$ on a Cauchy slice is conserved, i.e. it is  independent of the choice of the slice. Because of linearity, also the gravitational and matter currents individually are conserved. We want to find a function $H_{\xi}$ in phase space such that
\begin{align}
	\delta H_{\xi} =& - X_{\xi} \cdot \tilde{\Omega}, \label{eq:deltaH} \\
	\tilde{\Omega} :=& \int_{\Sigma} \tilde{\omega} .
\end{align}

Given a zero mode $\tilde{X}$ of the pre-symplectic form $\tilde{\Omega}$, we have
\begin{align}
	\tilde{X} \cdot \delta H_{\xi} &= \tilde{\Omega} \left( \tilde{X}, X_{\xi} \right) = 0,
\end{align}

guaranteeing that $H_{\xi}$ is a well-defined function in phase space. To check that there is a function in the phase space satisfying Eq. \eqref{eq:deltaH} we contract
\begin{align}
	- X_{\xi} \cdot \tilde{\omega} &= \left. \left( - X_{\xi} \cdot \delta \Theta + X_{\xi} \cdot \delta d C \right) \right|_{\tilde{\mathcal{P}}} \nonumber \\
	&= \left. \left( - \delta_{\xi} \Theta + \delta \left( X_{\xi} \cdot \Theta \right) + \delta_{\xi} d C - \delta \left( X_{\xi} \cdot d C \right) \right) \right|_{\tilde{\mathcal{P}}} \nonumber \\
	&= \left. \left( - \xi \cdot d \Theta - d \left( \xi \cdot \Theta \right) + \delta_{\xi} d C + \delta \left( X_{\xi} \cdot \Theta - X_{\xi} \cdot dC  \right) \right) \right|_{\tilde{\mathcal{P}}} \nonumber \\
	&= \left. \left( - \xi \cdot \delta L + \xi \cdot \left(E_{a} \delta \phi^{a}\right) + \delta \left(X_{\xi} \cdot \Theta \right) \right) \right|_{\tilde{\mathcal{P}}} + d \left( \delta_{\xi} C - \delta \left(X_{\xi} \cdot C \right) - \xi \cdot \Theta \right)  \nonumber \\
	&= \delta J_{\xi} + d \left( \delta_{\xi} C - \delta \left( X_{\xi} \cdot C \right) - \xi \cdot \Theta \right) \label{eq:Xomega}
\end{align}

After integrating this quantity on a Cauchy slice, we get
\begin{align}
	- X_{\xi} \cdot \tilde{\Omega} &= \int_{\Sigma} \delta J_{\xi} + \int_{\partial \Sigma} \left( \xi \cdot d C + d \left( \xi \cdot C \right) - \delta \left(X_{\xi} \cdot C \right) - \xi \cdot \Theta \right) \nonumber \\
	&= \delta \left( \int_{\Sigma} J_{\xi} + \int_{\partial \Sigma} \xi \cdot \ell - X_{\xi} \cdot C  \right),
\end{align}

where $d \left( \xi \cdot C \right) = 0$ because we take $\partial \Sigma$ to be a compact region. Hence
\begin{align}
	H_{\xi} &= \int_{\Sigma} J_{\xi} + \int_{\partial \Sigma} \xi \cdot \ell - X_{\xi} \cdot C .
\end{align}

For Weyl-transverse gravity coupled with matter we get
\begin{align}
	H_{\xi} &= - \frac{\Lambda - \lambda}{8 \pi G} \int_{\Sigma}  \xi \cdot \bm{\omega} + \int_{\partial \Sigma} Q_{\mathrm{GR}} + \xi \cdot \ell - X_{\xi} \cdot C + \text{constant} .
\end{align}

Using
\begin{align}
	\left. \xi \cdot \ell_{b} \right|_{\partial \Sigma} &= - \frac{b}{16 \pi G} \tilde{K} \xi^{\mu} \tilde{\tau}_{\mu} \bm{\omega}_{\partial \Sigma} , \\
	\left. X_{\xi} \cdot C_{G} \right|_{\partial \Sigma} &= \frac{\bm{\omega}_{\partial \Sigma}}{16 \pi G} \tilde{q}^{\mu\nu} \tilde{n}^{\alpha} \delta_{\xi} \tilde{g}_{\nu\alpha} \tilde{\tau}_{\mu} \nonumber \\
	&= \frac{\bm{\omega}_{\partial \Sigma}}{16 \pi G} \tilde{n}^{\alpha} \tilde{\tau}^{\nu} \left(\tilde{\nabla}_{\nu} \xi_{\alpha} + \tilde{\nabla}_{\alpha} \xi_{\nu} \right) \nonumber \\
	&= \frac{\bm{\omega}_{\partial \Sigma}}{16 \pi G} \left(\tilde{\tau}^{\alpha} \tilde{n}^{\beta} + \tilde{\tau}^{\beta} \tilde{n}^{\alpha} \right) \tilde{\nabla}_{\alpha} \xi_{\beta} , \\
	\left. Q_{\mathrm{GR}} \right|_{\partial \Sigma} &= - \frac{\bm{\omega}_{\partial \Sigma}}{16 \pi G} \left( \tilde{\tau}^{\alpha} \tilde{n}^{\beta} - \tilde{\tau}^{\beta} \tilde{n}^{\alpha} \right) \tilde{\nabla}_{\alpha} \xi_{\beta} ,
\end{align}

this leads to the Hamiltonian
\begin{align}
	H_{\xi} :=& - \frac{1}{8 \pi G} \int_{\partial \Sigma} \left( \tilde{\tau}^{\alpha} \tilde{n}^{\beta} \tilde{\nabla}_{\alpha} \xi_{\beta} + \frac{b}{2} \tilde{K} \xi^{\alpha} \tilde{\tau}_{\alpha} \right) \bm{\omega}_{\partial \Sigma} - \frac{\Lambda - \lambda}{8 \pi G} \int_{\Sigma} \omega_{\Sigma} \xi^{\mu} \tilde{\tau}_{\mu} + \int_{\partial \Sigma} \xi \cdot \ell_{\psi} - X_{\xi} \cdot C_{\psi} \nonumber \\
	&= - \frac{1}{8 \pi G} \int_{\partial \Sigma} \left( - \tilde{\tau}^{\alpha} \xi^{\beta} \tilde{\nabla}_{\alpha} \tilde{n}_{\beta} + \frac{b}{2} \tilde{K} \tilde{\tau}^{\alpha} \xi^{\beta} \tilde{q}_{\alpha\beta} \right) \bm{\omega}_{\partial \Sigma} - \frac{\Lambda - \lambda}{8 \pi G} \int_{\Sigma} \bm{\omega}_{\Sigma} \xi^{\mu} \tilde{\tau}_{\mu} + \int_{\partial \Sigma} \xi \cdot \ell_{\psi} - X_{\xi} \cdot C_{\psi} \nonumber \\
	&= - \frac{1}{8 \pi G} \int_{\partial \Sigma} \tilde{\tau}^{\alpha} \xi^{\beta} \left( - \tilde{K}_{\alpha\beta} + \frac{b}{2} \tilde{K} \tilde{q}_{\alpha\beta} \right) \bm{\omega}_{\partial \Sigma} - \frac{\Lambda - \lambda}{8 \pi G} \int_{\Sigma} \bm{\omega}_{\Sigma} \xi^{\mu} \tilde{\tau}_{\mu} \nonumber \\
	&\qquad \qquad \qquad + \int_{\partial \Sigma} \xi \cdot \ell_{\psi} - X_{\xi} \cdot C_{\psi} .
\end{align}

\subsection{First law of black-hole thermodynamics}

We can now examine how the Hamiltonian structure derived above modifies the first law of black-hole
thermodynamics. Consider a solution admitting a bifurcate Killing horizon generated by a Killing vector
field $\xi$. Since $X_{\xi}$ vanishes wherever $\xi$ is Killing, the Hamiltonian flow generated by $\xi$
trivializes in the bulk. Let $\Sigma$ be a Cauchy slice containing the bifurcation surface $\chi$, and let $\Sigma_{\mathrm{ext}}$
denote the portion of $\Sigma$ lying between $\chi$ and an external spatial boundary. Integrating
Eq.~\eqref{eq:Xomega} over $\Sigma_{\mathrm{ext}}$ yields
\begin{align}
	0 &= \int_{\Sigma_{\mathrm{ext}}} X_{\xi} \cdot \tilde{\omega} = \int_{\Sigma_{\mathrm{ext}}} \left( \delta J_{\xi} + d \left( \delta_{\xi} C - \delta \left( X_{\xi} \cdot C \right) \right) - \xi \cdot \Theta \right) \nonumber \\
	&= \delta H_{\xi}^{\mathrm{ext}} - \frac{\delta \Lambda}{8 \pi G} \int_{\Sigma_{\mathrm{ext}}} \xi \cdot \bm{\omega} -\int_{\chi} \delta Q_{\mathrm{GR}} - \xi \cdot \delta \ell + X_{\xi} \cdot C \nonumber \\
	&= \delta H_{\xi}^{\mathrm{ext}} - \frac{\delta \Lambda}{8 \pi G} \int_{\Sigma_{\mathrm{ext}}} \xi \cdot \bm{\omega}  - \int_{\chi} \delta Q_{\mathrm{GR}} , \label{eq:deltalambda1}
\end{align}

where we denote as $\delta H_{\xi}^{\mathrm{ext}}$ the contributions coming from the spacetime boundary, and we used the fact that $\xi$ vanishes on the bifurcation surface. As already noted in \cite{alonso-serrano_noether_2023}, the first law of thermodynamics now contains a term proportional to the variation of the cosmological constant. In general, $\delta\Lambda$ cannot be consistently set to zero without imposing additional restrictions on the space of solutions. It is therefore natural to compare this result with the extended phase-space formulation of black-hole thermodynamics, where $\Lambda$ is treated as a thermodynamic variable interpreted as pressure and the first law acquires a $V\,\delta P$ contribution (see, e.g., \cite{kubiznak_black_2017} and references therein). 
In WTG, the appearance of the $\delta\Lambda$ term is not postulated but follows
directly from the variational structure of the theory and the presence of the background volume form.
Moreover, the quantity conjugate to $\Lambda$ admits a natural geometric definition. For stationary
black-hole spacetimes with angular momentum, the first law takes the form
\begin{align}\label{eq:firstlaw}
	\delta M
	= \frac{\kappa}{8 \pi G}\,\delta A
	+ \Omega\,\delta J
	+ \frac{V}{8 \pi G}\,\delta \Lambda ,
\end{align}
where
\begin{align}
	V := \int_{\Sigma_{\mathrm{ext}}} \xi \cdot \bm{\omega} .
\end{align}

In contrast with standard black-hole chemistry, where several inequivalent notions of thermodynamic
volume exist, the definition of $V$ in WTG follows unambiguously from the
Hamiltonian analysis.

\newpage
%
\section{Conclusions}
%

In this work we developed the covariant phase space formulation of Weyl-transverse gravity  in the presence of timelike and spacelike boundaries. Although WTG reproduces the classical Einstein equations, its reduced gauge symmetry and the presence of a fixed background volume form introduce nontrivial modifications in the structure of the variational principle and the definition of canonical data. Our analysis makes these differences explicit.

Starting from the Weyl-invariant auxiliary metric, we derived the symplectic potential and presymplectic current while keeping all boundary contributions under control. This allowed us to identify the full set of boundary conditions compatible with a stationary action. The distinction between boundary conditions imposed on the auxiliary metric and those imposed on the dynamical metric becomes important in WTG, and it leads to a clean classification of Dirichlet and Neumann types for both objects. A notable outcome is that York boundary conditions follow naturally from the conformal structure of the theory wher the dynamical metric is precisely the quantity whose conformal class is fixed, making WTG an intrinsically adapted framework for such boundary data.

We constructed the Noether current and Hamiltonian generator for transverse diffeomorphisms and clarified how the Lagrangian ambiguity associated with the background volume form affects the definition of charges. The resulting surface term reproduces the usual GR expression supplemented by a contribution that depends on the integration constant arising in the Einstein equations. When evaluated on a bifurcate Killing horizon, the Hamiltonian identity leads to a first-law relation in which variations of the cosmological constant appear unless additional physical restrictions are imposed.
In this context, it is natural to compare our result with the extended phase-space thermodynamics program, where $\Lambda$ is promoted to a thermodynamic variable interpreted as pressure and the first law acquires a $V\delta P$ contribution (see, e.g., \cite{kubiznak_black_2017}, and references therein). Our derivation provides a clean covariant phase-space perspective: in WTG the appearance of a $\delta\Lambda$ term is not introduced by hand but arises directly from the variational structure of the theory and the presence of the background volume form. Whether this contribution should be retained or suppressed depends on global properties of the solution and on the choice of boundary conditions, rather than on thermodynamic assumptions. This offers a complementary viewpoint on the relation between black-hole chemistry and unimodular-like formulations of gravity. In fact, existing work in unimodular gravity considers the possibility of dark energy resulting from the dissipation of blackholes' angular momentum \cite{josset_dark_2017,perez_dark_2019}. A comparative work on the phenomenological results of the two fields is a viable path for future research.

Altogether, our construction supplies the complete boundary and phase-space framework required to study conserved quantities in WTG. It also isolates precisely which aspects of the theory depart from GR and which remain unchanged. The results open several natural extensions, including the analysis of null boundaries and corners, the characterization of asymptotic symmetries in flat and (A)dS regimes, and the study of whether the conformal structure of WTG leads to modified infrared behavior or memory effects. These directions can now be pursued within a consistent and fully covariant phase-space setting.

%
\section{Acknowledgements}
%

The authors thank Ana Alonso-Serrano, Marek Liška, Alejandro Perez, Yongge Ma, and Cong Zhang for fruitful discussions. SR is grateful for support from the National Natural Science Foundation of China (Grant No.12275022). G.O. is grateful for support from GA\v{C}R 23-07457S and GA\v{C}R 22-14791S grants of the Czech Science Foundation.

\bibliography{biblio.bib}


\end{document}